\documentclass[preprint,showpacs,superscriptaddress,preprintnumbers,amsmath,amssymb]{revtex4}
 
\usepackage{graphicx}
\usepackage{dcolumn}
\usepackage{bm}
 
\begin{document}
 
\preprint{Preprint}
 
\title{Proton decay from the isoscalar giant dipole resonance in $^{58}$Ni}
 
\author{M.~Hunyadi}
\affiliation{University of Notre Dame, Notre Dame, IN 46556, USA}
\affiliation{Institute of Nuclear Research of the Hungarian Academy of
      Sciences, H-4001 Debrecen, Hungary}	
	
\author{H.~Hashimoto}	
\affiliation{Research Center for Nuclear Physics, Osaka University,
	Mihogaoka 10-1, Ibaraki, 567-0047 Osaka, Japan}

\author{T.~Li}	
\affiliation{University of Notre Dame, Notre Dame, IN 46556, USA}

\author{H.~Akimune}
\affiliation{Department of Physics, Konan University, Kobe 658-8501, Japan}

\author{H.~Fujimura}	
\author{M.~Fujiwara}	
\affiliation{Research Center for Nuclear Physics, Osaka University,
Mihogaoka 10-1, Ibaraki, 567-0047 Osaka, Japan}

\author{Z.~G\'acsi}
\affiliation{Institute of Nuclear Research of the Hungarian Academy of
      Sciences, H-4001 Debrecen, Hungary}

\author{U.~Garg}
\affiliation{University of Notre Dame, Notre Dame, IN 46556, USA}

\author{K.~Hara}
\affiliation{Research Center for Nuclear Physics, Osaka University,
Mihogaoka 10-1, Ibaraki, 567-0047 Osaka, Japan}			

\author{M.N.~Harakeh}		
\affiliation{Kernfysisch Versneller Instituut, University of Groningen, 
9747 AA Groningen, The Netherlands}
\affiliation{GANIL, CEA/DSM â�� CNRS/IN2P3, F-14076 Caen, France}

\author{J.~Hoffman}		
\affiliation{University of Notre Dame, Notre Dame, IN 46556, USA}
 
\author{M.~Itoh}	
\affiliation{Research Center for Nuclear Physics, Osaka University,
Mihogaoka 10-1, Ibaraki, 567-0047 Osaka, Japan}

\author{T.~Murakami}	
\affiliation{Department of Physics, Kyoto University, Kyoto 606-8502, Japan}

\author{K.~Nakanishi}		
\affiliation{Research Center for Nuclear Physics, Osaka University,
Mihogaoka 10-1, Ibaraki, 567-0047 Osaka, Japan}

\author{B.K.~Nayak}	
\affiliation{University of Notre Dame, Notre Dame, IN 46556, USA}	

\author{S.~Okumura}
\affiliation{Research Center for Nuclear Physics, Osaka University,
Mihogaoka 10-1, Ibaraki, 567-0047 Osaka, Japan}

\author{H.~Sakaguchi}
\author{S.~Terashima}
\author{M.~Uchida}
\author{Y.~Yasuda}
\author{M.~Yosoi}
\affiliation{Department of Physics, Kyoto University, Kyoto 606-8502, Japan}

\date{\today}

\begin{abstract}

Proton decay from the 3$\hbar\omega$ isoscalar giant 
dipole resonance (ISGDR) in $^{58}$Ni has been measured using the ($\alpha,\alpha'p$) reaction 
at a bombarding energy of 386 MeV to investigate its decay properties.
We have extracted
the ISGDR strength under the coincidence condition between inelastically scattered $\alpha$ particles at forward
angles and decay protons emitted at backward angles.
Branching ratios for proton decay to low-lying states of $^{57}$Co have been determined, and the results
compared to predictions of recent continuum-RPA calculations.
The final-state spectra of protons decaying to the low-lying states in $^{57}$Co were 
analyzed for a more detailed understanding of the structure of the ISGDR.
It is found that there are differences in the structure of the ISGDR as a function of excitation energy.

\end{abstract}
 
\maketitle
 
\section{\label{sec:introduction} INTRODUCTION}
 
Experimental studies of direct decay modes of giant resonances can shed light on
their microscopic structure and its evolution to a collective behavior.
Since most of the giant resonances are located above the particle separation
thresholds, particle emission is the dominant decay mode that can take place
at different stages of the damping process \cite{ha01}.
Particle decay can occur either from the initial 1p-1h state leaving a
single-hole state in the $A-1$ nucleus, or from the states
with partially or completely equilibrated configurations,
resulting in an evaporation-like spectrum.
Therefore, the branching ratios of particle emission from giant resonances
provide us a good means to understand the wave function of the giant resonance.

The main interest in studying isoscalar giant resonances is to investigate the 
nuclear incompressibility via the determination of the resonance energies of the isoscalar giant
monopole resonance (ISGMR) and the isoscalar giant dipole resonance (ISGDR)
\cite{ha81,st82,co00,sh02}.
The incompressibility has a crucial role in the application of the
nuclear equation-of-state in model calculations of various astrophysical
processes or in descriptions of heavy-ion reactions.
These experiments had focused primarily on the ISGMR as the strongest compression resonance ever
discovered \cite{ha77,yo77}, while the ISGDR could be studied in detail only after
the first unambiguous evidence for its existence in 1997 \cite{da97}.
Difficulties were mainly caused by the relatively weak excitation of
the 3$\hbar\omega$ ISGDR, which, in addition, was distributed
with a relatively large width on top of a strong nuclear-continuum and instrumental background.
An overlap in excitation energy with the 3$\hbar\omega$ high-energy octupole resonance (HEOR)
had to be separated from the ISGDR strength as well.

The problem of separating the background due to the nuclear continuum from
the resonance strengths still
remained due to the unknown multipole composition.
To identify the multipole components, we employ the multipole-decomposition analysis (MDA)
to fit the angular distribution of the obtained cross sections, or the difference-of-spectra (DOS)
technique, in which pronounced changes and large differences
in the shapes of the angular distributions for low multipoles are exploited in measurements
at very forward angles, possibly including 0$^\circ$ as well \cite{br87,da97}.
The elimination of the background components was the major difficulty in many of the singles experiments.
There remained some ambiguities regarding an appropriate choice of the background-subtraction
resulting in a large uncertainty in the shape and the strength of giant resonances.
Moreover, measurements at scattering angles close to 0$^\circ$ often suffered from a
substantial contribution of instrumental background in addition to the non-resonant continuum,
which is due to quasi-free scattering and pick-up/break-up reactions.
The coincidence technique with particle decay gives a useful means to suppress such backgrounds
and to isolate resonances strengths.
The ISGDR strength can be identified using the DOS analysis provided data are measured in
appropriately chosen angular regions where dipole strength dominates over those of
other multipolarities (see Fig.~\ref{fig:dwba}).

In previous experiments, coincidence measurements with decay particles from high-energy resonances 
and the subsequent DOS analysis were found to be effective for eliminating both the instrumental
background and the non-resonant continuum \cite{hu03,hu04,hu07}.
The resulting resonance strength and its partial decay branches
to single-hole states give detailed information on the microscopic structures of these resonances.

Experiments on the ISGDR in $^{58}$Ni, with a closed shell of 28 protons, performed so far were
exclusively singles measurements in which $L$=1 strength distributions were analyzed
with MDA techniques over a wide excitation energy range \cite{lu00,lu06,na06}.
A fraction of the $L$=1 strength was systematically observed at E$_{x}$ $\sim $ 16 MeV
(low-energy part), but the main part of the ISGDR at E$_{x}$ $\sim $ 31 MeV was only found
in the latest experiments \cite{lu00,na06}.
Coincidence experiments for the ISGDR have been performed so far only for $^{90}$Zr, $^{116}$Sn
and $^{208}$Pb, where gross resonance parameters (energies, widths) and partial branching 
ratios were determined \cite{hu03,hu04,hu07,na09}.
We have extended the coincidence studies towards lighter nuclei
and investigated the microscopic structures based on the observation of an increasing
fragmentation of resonance strengths and decay properties.

Here, we report the first results of a coincidence experiment,
in which we aimed at studying the $L$=1 strength distribution in the 3$\hbar\omega$
region of $^{58}$Ni, by applying the analysis with the DOS technique, and at deducing
its direct proton-decay properties to low-lying proton-hole states of $^{57}$Co.

\section{\label{sec:experiment} EXPERIMENT}
 
The experiment was carried out at the Ring cyclotron facility of the Research Center for Nuclear
Physics (RCNP) in Osaka University, using the $^{58}$Ni($\alpha,\alpha'p$) reaction at a beam energy
of 386 MeV.
We used a $^{58}$Ni target with a thickness of 4 mg/cm$^2$.
Inelastically scattered $\alpha$-particles were momentum-analyzed at very forward angles
with the high-resolution magnetic spectrometer, Grand Raiden \cite{fu99}.
The focal-plane detector system of the spectrometer consisted of two multi-wire 
drift chambers (MWDC) and two plastic scintillators.
The spectrometer enables the detection of ejectiles with a momentum acceptance of $\delta p/p$=5\%
corresponding to an effective excitation-energy range of 15-50 MeV.
The spectrometer was set at the scattering angle of $\Theta_{lab}$=2.5$^\circ$, 
with an opening angle of 40 mr both horizontally and vertically defined by the 
geometry of the entrance slit, which corresponds to an effective angular 
acceptance of 1.8$^\circ$-3.8$^\circ$.

Decay protons were analyzed with a detector array consisting of
13 Si(Li) detectors with a thickness of 5 mm each.
These were mounted in the backward hemisphere in the scattering chamber at a radial
distance of 12 cm from the target.
A total solid angle of 1.6 msr in the range of 
100$^\circ$-220$^\circ$ in the laboratory frame was covered.
The resolutions in the $^{58}$Ni($\alpha,\alpha'$) spectrum and in the $^{57}$Co
final-state spectrum were 200 keV and 400 keV full width at half maximum (FWHM), respectively.

\section{DISCUSSION}

Energy spectra and angular distributions of inelastically scattered $\alpha$ particles
were measured both in singles and in coincidence with protons emitted from the highly excited
states in $^{58}$Ni.
Spectra of final-state energy, defined as the excitation energy of the $^{57}$Co daughter
nucleus, were derived from the excitation energies in $^{58}$Ni and proton decay energies.

\begin{figure}[t]
\includegraphics{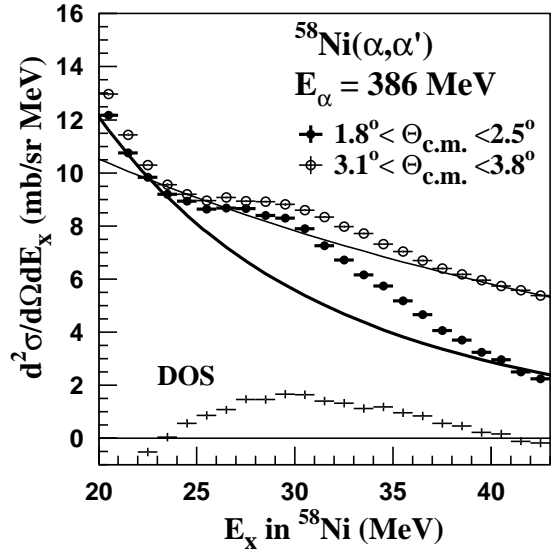}
\caption{\label{fig:singlesexc} Inelastic $\alpha$-scattering spectra at 
$1.8^\circ<\Theta_{c.m.}<2.5^\circ$ and $3.1^\circ<\Theta_{c.m.}<3.8^\circ$. 
The resonance structures were separated from the background with functions 
fitted in the continuum regions at $E_x$=23-40 MeV (see text for more details); 
the subtracted backgrounds are shown by the solid lines.
The difference of the two spectra (DOS) after subtracting the assumed backgrounds is also shown.}
\end{figure}

Singles ($\alpha,\alpha'$) spectra were analyzed to obtain differential
cross sections for the ISGDR strength to determine the fraction of the energy-weighted sum
rule (EWSR) exhausted.
The ($\alpha,\alpha'$) spectra obtained with the opening angle of 40 mr both horizontally
and vertically, corresponding to scattering angles $\Theta_{c.m.}=1.8^\circ-3.8^\circ$,
were subdivided into two regions of $\Theta_{c.m.}=1.8^\circ-2.5^\circ$ and 
$3.1^\circ-3.8^\circ$ by using the ray-tracing technique.
These two spectra are given in Fig. \ref{fig:singlesexc}, both of which show a resonance
of the ISGDR in the 3$\hbar\omega$ region on top of a nuclear-continuum background.
In these spectra the contribution from the instrumental background, which was mainly due to
scattering events of the incident beam on the slits defining the opening angle and on various
parts of the spectrometer, was eliminated by subtraction.
This was possible because it has been observed that these background events at the focal plane
had a flat distribution in the vertical position spectrum, while true events were focused
at the focal plane both horizontally and vertically, which allowed their separation
from the instrumental background \cite{uc03}.

\begin{figure}[t]
\includegraphics{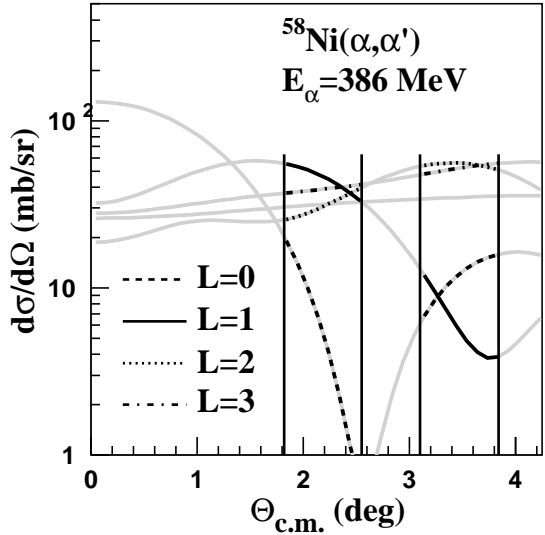}
\caption{\label{fig:dwba} Differential cross sections of the $^{58}$Ni($\alpha,\alpha'$) reaction
at $E_\alpha$=386 MeV for exciting isoscalar giant resonances in $^{58}$Ni.
The distorted-wave Born approximation (DWBA) calculations have been obtained by assuming 100\% exhaustion
of the energy-weighted sum rules (EWSR). 
The resonance energies have been assumed to be at $E_x$=19, 29, 15 and 26 MeV for the
$L$=0, 1, 2 and 3 resonances, respectively.
The regions used for the difference-of-spectra (DOS) analysis are indicated.}
\end{figure}

A remarkable difference between the spectra in Fig. \ref{fig:singlesexc} is due to the presence
of the $L$=1 excitation strength as expected (see Fig. \ref{fig:dwba}).
We also observed that the nuclear-continuum background changed in shape and amplitude.
Since the nuclear continuum comprises, in general, a mixture of many multipoles,
such changes in the background could be treated by making reasonable assumptions on its shape.
To define a background under the resonance, its shape can typically be approximated by simple functions
such as polynomials that interpolate the nuclear continuum between the continuum above the 
3$\hbar\omega$ region, where significant contributions of 4$\hbar\omega$ resonances are not expected 
($E_x>40$ MeV), and an intermediate region between the ISGDR and ISGMR (around 23 MeV), 
where the contributions from the high-energy tail of the latter are assumed to be negligibly small.
Various background functions and fitting conditions were tested to
give a reliable estimate for the systematic errors of integrated ISGDR cross sections
with the $L$=1 strengths at $E_x=$23-40 MeV.
As a result, we obtained the cross section values of 24.1$\pm$1.1 mb/sr at 
$1.8^\circ<\Theta_{c.m.}<2.5^\circ$ and 7.1$\pm$1.0 mb/sr for $3.1^\circ<\Theta_{c.m.}<3.8^\circ$.
Since the non-resonant continuum background underlying the ISGDR was assumed to be completely removed 
using different hypothetical functions, only the singles spectrum resulting from the difference-of-spectra 
(DOS) was plotted for comparison with the spectra obtained with the gate on decay proton 
events in order to deduce the branching ratios.

\begin{figure}[t]
\includegraphics{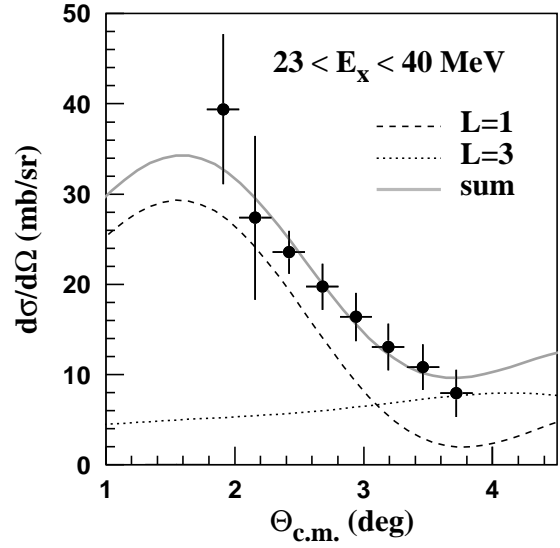}
\caption{\label{fig:singlesad} Angular distribution of the cross sections for the resonance 
structure observed in the 3$\hbar\omega$ region. The cross sections for the resonance at
$E_x$=23-40 MeV were fitted with the DWBA calculations for $L$=1 and $L$=3.
The $L$=1 and $L$=3 cross sections used in the fit are plotted.}
\end{figure}

The decrease of the DOS below E$_x$=20 MeV is understood as a negative 
contribution of the ISGMR and ISGQR, as indicated from the angular distributions shown in 
Fig. \ref{fig:dwba}.
A more detailed analysis of singles data was accomplished by multipole decomposition
of the resonance structures in the 3$\hbar\omega$ region after subtraction of the
nuclear-continuum background.
Angular distributions were determined by integrating double-differential cross sections 
at $E_x$=23-40 MeV after subtracting the background functions
shown in Fig. \ref{fig:singlesexc}.
The cross sections of the 3$\hbar\omega$
resonance structure are compared with the results from the distorted wave Born approximation
(DWBA) calculations, and are fitted by using $L$=1 and $L$=3 components.
The result is shown in Fig. \ref{fig:singlesad}.
It is found that the resonance bump at E$_x$=23-40 MeV consists of mainly the $L$=1 ISGDR component
and a small contribution of the 3$\hbar\omega$ HEOR, at least, at the scattering angles of 
$\Theta_{c.m.}$=1.8$^{\circ}-3.8^\circ$.
Any contributions from the $L$=0 and $L$=2 resonances are not evidenced in the fit.

The analysis of coincidence data started with the identification of prompt and random
coincidence events in the timing spectra.
Since the timing resolution of 1.9 ns was small enough relative to the repetition period of 19.7 ns
for beam bursts, prompt events were well isolated and could be corrected for random
coincidences.
The real-to-random ratio was found to be $\approx$ 10 after excluding $\gamma$-ray events
detected by the Si(Li) detectors.
Moreover, since non-resonant quasi-free scattering of protons is strongly
forward peaked, its contribution to decay spectra at backward angles, where
all proton detectors were placed, was negligibly small.

Angular correlations of decay protons and the $\alpha$-particle ejectiles were also extracted
and fitted with Legendre polynomials to calculate the yields of proton emission,
which were necessary to determine the absolute values of cross sections and branching ratios
for proton decay.

\begin{figure*}[t]
\includegraphics{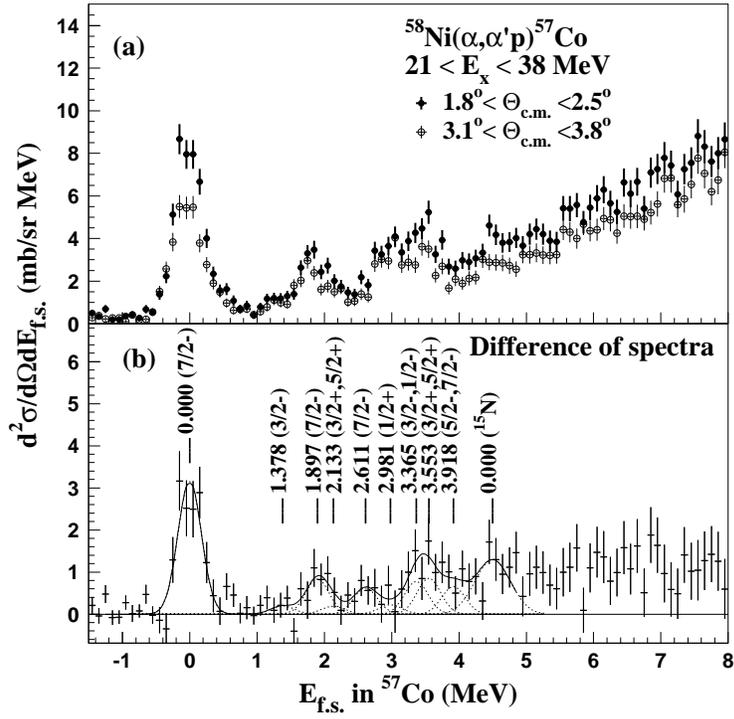}
\caption{\label{fig:efsdos} Double-differential cross sections as function of the final-state
energy in $^{57}$Co projected for the 3$\hbar\omega$ region in $^{58}$Ni.
a) Final-state spectra extracted for the scattering angles $1.8^\circ<\Theta_{c.m.}<2.5^\circ$ and 
$3.1^\circ<\Theta_{c.m.}<3.8^\circ$.
b) DOS showing the feeding pattern to the low-lying states in $^{57}$Co by emitting protons
from the ISGDR. 
Individual final states are partly identified and fitted.
Some yield corresponding to the $^{16}$O($\alpha,\alpha'p)^{15}$N$_{g.s.}$ reaction
is also observed due to oxygen contamination in the target.}

\end{figure*}

Final-state spectra as a function of excitation energy of the daughter nucleus
$^{57}$Co were extracted for the scattering angles 1.8$^\circ<\Theta_{c.m.}<2.5^\circ$ and
3.1$^\circ<\Theta_{c.m.}<3.8^\circ$, and for the excitation-energy range 21-40 MeV where
the ISGDR was observed in the singles $^{58}$Ni($\alpha,\alpha'$) spectra.
The final-state spectra and the spectrum resulting from their subtraction are shown
in Fig. \ref{fig:efsdos}.
The prominent peak due to proton decay from the ISGDR to the $f_{7/2}$ proton-hole ground
state of $^{57}$Co is clearly observed and is well separated from the other
low-lying states.
It is evident that the ISGDR decays to the $f_{7/2}$ proton-hole state in $^{57}$Co
by emitting protons due to the 1p-1h configuration in its wave function.
Most of the excited states up to $E_x$=3 MeV in $^{57}$Co are identified in the final-state 
spectrum.
Due to the energy resolution of 400 keV in the final-state spectrum and due to the increasing
level density as a function of excitation energy, the low-lying states at $E_x\geq$1 MeV could
not be well resolved.
It is worth noting that we observed a weak yield of the ground-state decay from the $^{16}$O
to $^{15}$N due to oxygen contamination in the target material.
However, we conclude that there is no other background contribution at $E_x<4.5$ MeV.

Event yields in both spectra in Fig. \ref{fig:efsdos}a show a monotonous increase at $E_x>4$ MeV,
which is attributed to the statistical component of decay protons from both the ISGDR and 
non-resonant continuum to the states in $^{57}$Co with multi-particle-multi-hole configurations.
Statistical decay calculations using the code CASCADE were performed and it was found that the final-state spectra
at $E_x>4$ MeV are well described by the statistical-model calculations.
On the other hand, the calculation of proton decay for the low-lying final
states becomes strongly model-dependent; still, it is concluded from the statistical decay calculations that
proton decay to the ground state in $^{57}$Co proceeds dominantly through the direct process.

\begin{figure*}[t]
\includegraphics{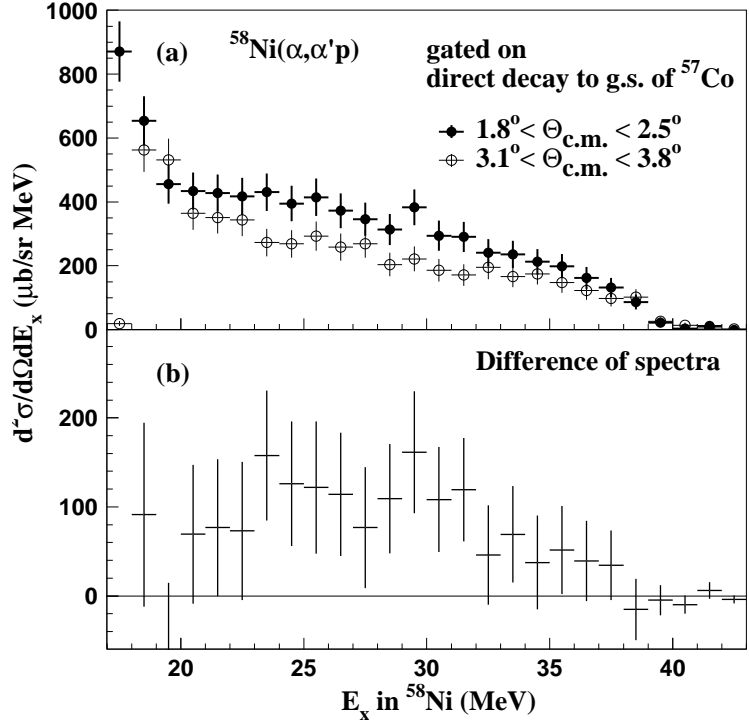}
\caption{\label{fig:excdos} a) Double-differential cross sections at 
$1.8^\circ<\Theta_{c.m.}<2.5^\circ$ and $3.1^\circ<\Theta_{c.m.}<3.8^\circ$ plotted
as a function of excitation energy in $^{58}$Ni.
The data points have been obtained with a gate on the proton decay to the ground state in $^{57}$Co. 
b) DOS representing the $L$=1 strength populating the ground state in $^{57}$Co.}
\end{figure*}

The ground state of $^{57}$Co has a simple wave function with a single $f_{7/2}$ proton-hole configuration. 
This state is readily fed from the resonance with a particle-hole configuration in the proton orbitals. 
Since knock-out and pick-up/break-up and break-up/pick-up processes lead to very forward emission of protons, 
measuring proton decay at backward angles ($>90^\circ$) eliminates the contribution of these processes. 
The remaining underlying non-resonant continuum background due to multistep processes will display 
a dominant statistical decay with no direct decay component, and a flat angular distribution is expected 
for this continuum at very forward angles. 
By gating on decay to the $f_{7/2}$ proton-hole ground state, 
which does not comprise statistical decay components according to the CASCADE calculations discussed above, 
resonant structures are selected in the 3$\hbar\omega$ region, including the ISGDR, and non-resonant 
contributions are suppressed from the excitation-energy spectra obtained by the DOS technique. 
Figure \ref{fig:excdos}a shows the excitation energy spectra in $^{58}$Ni gated on the proton 
decay to the ground state in $^{57}$Co for the angles
1.8$^\circ < \Theta_{c.m.} < 2.5^\circ$ and 3.1$^\circ < \Theta_{c.m.} < 3.8^\circ$. 
A significant difference is observed for the spectrum taken at
1.8$^\circ < \Theta_{c.m.} < 2.5^\circ$ compared to the spectrum taken at 3.1$^\circ < \Theta_{c.m.} < 3.8^\circ$. 
As seen in Fig. \ref{fig:excdos}b, a prominent structure appears after their subtraction in an 
excitation-energy range of $E_x$=20-40 MeV, where a major fraction of the ISGDR strength exists. 
In the excitation energy region at the high-energy side of the ISGDR, the background of the nuclear 
continuum is evidently suppressed, which justifies the validity of the analysis combined with 
both the DOS technique and the coincidence technique for the observation of the direct particle 
decay channels.

\begin{figure*}[t]
\includegraphics{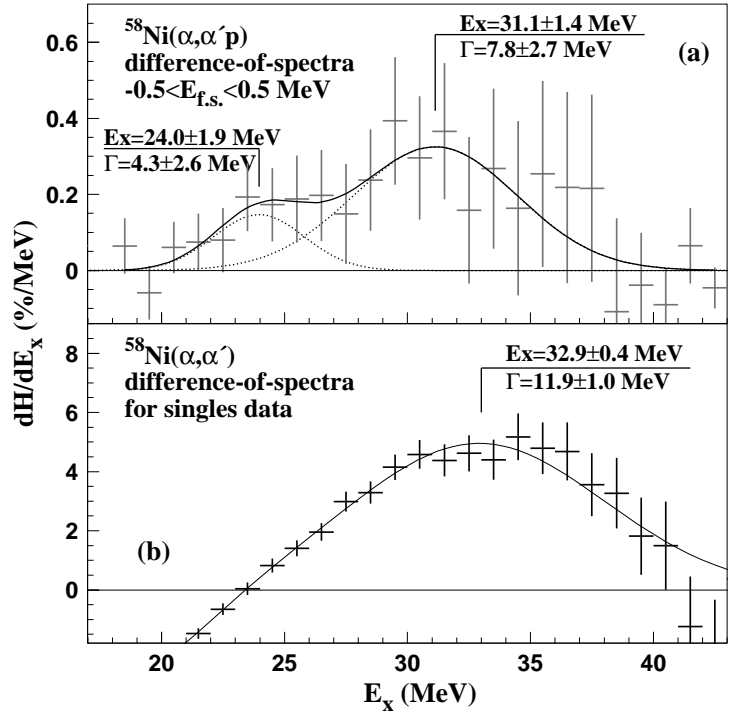}
\caption{\label{fig:strength} Strength distributions expressed in differential exhaustion
factors defined as fractions of the EWSR of ISGDR.
The spectra represent the $L$=1 strengths identified with DOS analysis of a) coincidence data 
gated with proton-decay to the ground state in $^{57}$Co (Fig. \ref{fig:excdos}) and 
b) singles data (Fig. \ref{fig:singlesexc}). Fit functions and parameters are also indicated.}
\end{figure*}

The spectrum obtained with the DOS technique with a gate on proton decay to
the ground state in $^{57}$Co, shown in Fig. \ref{fig:excdos}b, was converted to
a strength distribution based on DWBA calculations, and was expressed
in terms of the fraction of the EWSR ($dH/dE_x$).
The resulting strength distribution is shown in Fig. \ref{fig:strength}a.
Similarly, the singles cross section for the ISGDR obtained by the DOS method (labeled as DOS in Fig. \ref{fig:singlesexc}) was converted to a strength distribution and expressed as fraction of EWSR rule. 
This is shown in Fig. \ref{fig:strength}b.
The resonance bump of the ISGDR shown in Fig. \ref{fig:strength}a
now seems to consist of, at least, two components.
Firstly, the resonance shape was tested by fitting with a simple Gaussian function covering
the whole resonance region.
This procedure was not successful in getting a good fit. 
Secondly, we tried to fit the resonance using two Gaussian functions 
for the low- and high-energy parts.
This leads to a reduced $\chi^2$ value of 0.94.
The resonance parameters for the fit are listed in Table \ref{tab:tabI}.

The centroid energy and moments were also derived for the partial strength distribution obtained for
the direct proton decay channel to the ground state of $^{57}$Co (see Table I). However, the determination of the strength distributions for all possible decay channels would be essential to obtain
the centroid energy and moments of the ISGDR and, consequently, the bulk resonance properties, as
well as the incompressibility from centroid energies. Contributions from the direct alpha-decay channels--the alpha-separation threshold is very low (S$_{\alpha}$ = 6.400 MeV)--were, nevertheless, found to be negligible in the excitation
energy region of the ISGDR and did not yield significant information on its partial strength distribution. Considerable yields from direct neutron decay comparable to those from proton decay are still expected, since the neutron decay threshold is only around 4
MeV higher than the proton threshold (S$_{n}$=12.217 MeV, S$_{p}$=8.172 MeV). Therefore, it would be very interesting to extend these studies to measuring neutron decay of the ISGDR.

In medium-heavy nuclei, a fragmented resonance structure reminiscent of the underlying 
1p-1h structure is predicted by RPA calculations (see Ref. \cite{co04}).
This kind of fragmentation could be observed in the coincidence measurements with decay protons,
if we can obtain the final-state spectra with considerable statistics
and with a resolution good enough to resolve the low-lying final states in $^{57}$Co.
It should be noted that in the present experiment, these requirements were not 
met to obtain the strength distributions of decay protons to the excited final states in $^{57}$Co.
Therefore, it becomes difficult to obtain further information on the ISGDR structure from further 
analyses of the final-state spectrum.
In spite of these limitations, it is quite apparent that the ISGDR consists of, at least, 
two components, with the low- and high-energy parts located at E$_x$=21-27 MeV and E$_x$=27-38 MeV, respectively.

\begin{figure*}[t]
\includegraphics{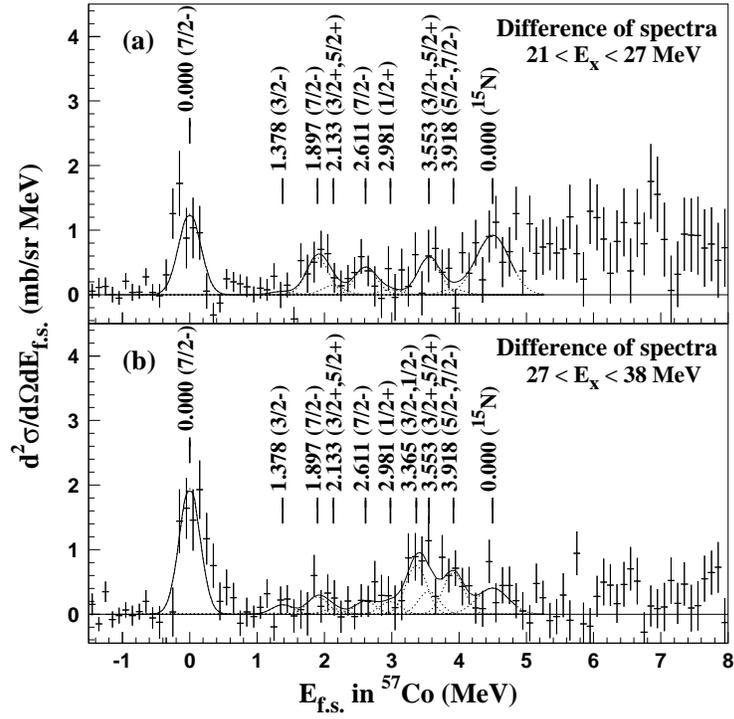}
\caption{\label{fig:dosboth} Difference final-state spectra plotted as function of the excitation
energy in $^{57}$Co. Coincidence conditions are set for a) low- and b) high-energy 
components of the ISGDR strength in $^{58}$Ni. 
Fits of individual final states are plotted.}
\end{figure*}

Figure \ref{fig:dosboth} shows the difference final-state spectra generated by subtracting 
the final-state spectra taken at 1.8$^\circ < \Theta_{c.m.} < 2.5^\circ$ and at
3.1$^\circ < \Theta_{c.m.}<3.8^\circ$ and gated on proton decay from the two ISGDR 
components in $^{58}$Ni. Therefore, these final-state spectra correspond to proton decay 
from these two ISGDR components. 
Some peaks corresponding to the ground state and the low-lying states of $^{57}$Co were fitted 
with Gaussians.
Many other peaks had to be left out in the fitting procedure because of poor statistics 
and the relatively high level densities that did not allow to resolve 
individual low-lying states due to a resolution of 400 keV in the final-state spectrum.
However, significant differences are found between the spectra in Figs. \ref{fig:dosboth}a and \ref{fig:dosboth}b.
This can be understood by taking into account the fact that the
low-lying states in $^{57}$Co have relatively complex wave functions, since $^{58}$Ni has 28 protons 
and is in the middle of the $fp$-shell, and their structure can be described mostly as one-hole 
coupled to phonon excitations with further coupling to $n$-particle-($n$+1)-hole configurations.
A reliable interpretation of their population rates would thus require a further improvement
in both the experimental data (statistics and resolution) and RPA calculations.
If such high-statistics and high-resolution coincidence ($\alpha,\alpha'p$) 
experiment were successfully performed, one would obtain
rich information on the internal structure and evolution of the
particle-hole configuration of the ISGDR in order to understand correctly the damping process 
of the resonance.
To demonstrate the differences in proton decay from the two regions
of the ISGDR structure to the low-lying states in $^{57}$Co, 
their relative strength was deduced for each 1 MeV bin of
the final-state energy.
This quantity can be considered an asymmetry parameter of the ISGDR strength distribution,
and used for testing the theoretical models.
The results are listed in Table \ref{tab:tabII}.

Experimental branching ratios of proton decay to the ground state (g.s.) and to all states below 
$E_{f.s.}$=4 MeV (including the g.s.) were deduced to be 6.2$\pm$1.4\% and 12.9$\pm$2.3\%, respectively.
These were determined from the total ISGDR strengths in the coincidence and singles difference spectra in the excitation-energy region of $E_x$=23-38 MeV shown in Figs. \ref{fig:strength}a and b.
These values are quite considerable even in comparison with continuum-RPA calculations,
which predicted 16.4\% for the single-hole $f_{7/2}$ level taking a unit spectroscopic 
factor \cite{go04}.

In conclusion, we have reported on coincidence measurements to investigate the proton decay
from the ISGDR using the $^{58}$Ni$(\alpha,\alpha'p)$ reaction at $E_\alpha$=386 MeV.
The ISGDR has been observed at $E_x$=20-40 MeV and has been identified by employing
difference-of-spectra (DOS) analysis based on the singles and coincidence measurements.
The strength distributions of the ISGDR in terms of fractions of the EWSR have been deduced
and fitted to determine the gross structure of the ISGDR from both singles and proton-decay data.
The gross structure of the ISGDR determined from coincidence measurements shows two components.
Significant differences were found in proton decay from these two components of the ISGDR 
to the low-lying states in $^{57}$Co at excitation energies below 4 MeV.
This experimental result indicates that there exist some structural differences 
in the particle-hole configurations and damping process of the ISGDR in
$^{58}$Ni as a function of excitation energy.
Partial branching ratios for proton decay to the ground state and the low-lying final 
states up to 4 MeV in $^{57}$Co have been determined.
A deeper understanding of the branching ratios and decay patterns of proton
decay from the ISGDR would require further development of theoretical models.

\begin{acknowledgments}

We thank the cyclotron staff at RCNP for their support during the experiment
described in this paper.
M.H. thanks the University of Notre Dame for the hospitality
during his long stay for the research work.
This work has been supported in part by the U.S. National Science
Foundation (grant Nos. INT03-42942, PHY01-40324, PHY05-52843, and PHY07-58100)
and the US-Japan Cooperative Science Program of Japan Society
for the Promotion of Science (JSPS).
The Hungarian co-authors thanks the support of the Hungarian fund OTKA
under contract K72566.

\end{acknowledgments}

\bibliography{ni58}

\vfill\eject

\begin{table*}[t]
\caption{\label{tab:tabI}
Resonance parameters and fractions of the energy-weighted sum rule (EWSR) determined for the
ISGDR based on Gaussian fits 
and the moments of strength distributions shown in Figs. \ref{fig:strength}a and b.
Moments $m_1$ and $m_0$ were derived for the excitation-energy region of 23-38 MeV.
Theoretical values are taken from continuum-RPA calculations. 
The symbols $\omega, \Gamma$ and $H$ denote the centroid energy of the ISGDR, the width
and fraction of EWSR in \%, respectively, and $m_0$ and $m_1$ are moments of the strength
distribution defined by $m_k=\int E^kSds$ \cite{go04}.}
\begin{ruledtabular}
\begin{tabular}{lccccc}
			& singles data 	& singles data	& singles data	& coincidence data\footnote{data 
gated on direct decay to the g.s. of $^{57}$Co.}	& theory	\\
			& {\it this work}& Ref. \cite{lu06} & Ref. \cite{na06} & {\it this work} & Ref. \cite{go04}\\
\hline			& & & & \\
Gaussian fit		& & & & \\
$\omega$ (MeV)   	& 32.9$\pm$0.4	& 34.06$\pm$0.30& & 29.6$\pm$1.1	& 	\\
$\Gamma$ (MeV)		& 11.9$\pm$1.0	& 19.52$^{+0.41}_{-0.40}$& & 10.8$\pm$1.9	& 	\\
$H$ (\%)		& 63$\pm$5	& 		& & 3.4$\pm$0.6		& 	\\
			& & & & & \\
Double-Gaussian fit	& & & & & \\
$\omega$ (MeV) 		&		&		& & 24.0$\pm$1.9, 31.1$\pm$1.4	& \\
$\Gamma$ (MeV)		&		&		& & 4.3$\pm$2.6, 7.8$\pm$2.7	& \\
$H$ (\%)		&		&		& & 0.90$\pm$0.17, 2.5$\pm$0.4	& \\
			& & & & & \\
Moments			& & & & & \\
$m_1/m_0$ (MeV)   	& 29.5$\pm$2.0	& 27.78$^{+0.47}_{-0.30}$ & 30.8$^{+1.7}_{-1.1}$ & 27.9$\pm$7.5 & 27.4 \\
$m_1$ (\%)		& 54$\pm$8	& 68$^{+20}_{-15}$	& 98$^{+4}_{-5}$ & 3.4$\pm$0.4	& 75 	\\
\end{tabular}
\end{ruledtabular}
\end{table*}

\begin{table}[t]
\caption{\label{tab:tabII}
Ratio of the high-energy part ($E_x$=27-38 MeV) of the ISGDR to the total ($E_x$=21-38 MeV), 
derived as a function of final-state energy in $^{57}$Co.
The final-state energy has a 1 MeV bin size.}
\begin{ruledtabular}
\begin{tabular}{ccc}
$\langle E_{f.s.}\rangle$	& $H_{HE}/(H_{LE}+H_{HE})$ & $H_{LE}+H_{HE}$ \\
(MeV)		& (\%)	& (\% of EWSR) \\\hline
0 & 74$\pm$15		& 3.40$\pm$0.41 	\\
1 & n/a			& n/a 		\\
2 & 45$\pm$25		& 1.15$\pm$0.29	\\
3 & 81$\pm$22		& 1.96$\pm$0.35	\\
4 & 63$\pm$16		& 2.14$\pm$0.32	\\
5 & 34$\pm$18		& 1.67$\pm$0.34	\\
6 & 43$\pm$17		& 2.12$\pm$0.39	\\
7 & 30$\pm$18		& 1.95$\pm$0.42	\\
\end{tabular} 
\end{ruledtabular}
\end{table}

\end{document}